\documentclass[journal=jacsat,manuscript=article]{achemso}

\usepackage[none]{hyphenat}
\def\blue#1{{\color{blue}#1}}
\usepackage[pdftex,colorlinks=true,linkcolor=Cerulean,citecolor=Cerulean,urlcolor=Cerulean]{hyperref}
\hypersetup{linkcolor=blue,citecolor=blue,urlcolor=blue}
\hypersetup
{
	pdfauthor={
		J. Enrique V\'azquez\,-Lozano (juavazlo@ntc.upv.es) \&
		Alejandro Mart\'inez (amartinez@ntc.upv.es)},
	pdfsubject={
		Nanophotonics Technology Center [Universitat Polit\`ecnica de Val\`encia]},
	pdftitle={
		Towards Chiral Sensing and Spectroscopy Enabled by All-Dielectric Integrated Photonic Waveguides},
	pdfkeywords={optical chirality, circular dichroism, chiral sensing, chiroptical spectroscopy, integrated photonics, localized resonance}
}
\usepackage{chemformula} % Formula subscripts using \ch{}
\usepackage[T1]{fontenc} % Use modern font encodings

\author{J. Enrique V\'azquez-Lozano}
\affiliation{Nanophotonics Technology Center, Universitat Polit\`ecnica de Val\`encia, Camino de Vera s/n, 46022 Valencia, Spain}
\author{Alejandro Mart\'inez}
\affiliation{Nanophotonics Technology Center, Universitat Polit\`ecnica de Val\`encia, Camino de Vera s/n, 46022 Valencia, Spain}
\email{amartinez@ntc.upv.es}

\title{Towards Chiral Sensing and Spectroscopy Enabled by All-Dielectric Integrated Photonic Waveguides}

\abbreviations{CPL,TE,TM,NIR,CPGM,PEC,PBL}

\begin{document}
\maketitle
\sloppy


%\begin{tocentry}
%
%Some journals require a graphical entry for the Table of Contents.
%This should be laid out ``print ready'' so that the sizing of the
%Inside the \texttt{tocentry} environment, the font used is Helvetica
%8\,pt, as required by \emph{Journal of the American Chemical
%The surrounding frame is 9\,cm by 3.5\,cm, which is the maximum
%permitted for  \emph{Journal of the American Chemical Society}
%graphical table of content entries. The box will not resize if the
%content is too big: instead it will overflow the edge of the box.
%
%This box and the associated title will always be printed on a
%separate page at the end of the document.
%
%\end{tocentry}

\begin{abstract}
Chiral spectroscopy is a powerful technique that enables to identify the chirality of matter through optical means. So far, experiments to check the chirality of matter or nanostructures have been carried out using free-space propagating light beams. However, for the sake of miniaturization, it would be desirable to perform chiral spectroscopy in photonic integrated platforms, with the additional benefit of massive parallel detection, low-cost production, repeatability, and portability of such a chiroptical device. Here we show that all-dielectric integrated photonic waveguides can support chiral modes under proper combination of the fundamental eigenmodes. In particular, we investigate two mainstream configurations: a dielectric wire with square cross-section and a slotted waveguide. We analyze numerically three different scenarios in which such waveguides could be used for chiral detection: all-dielectric waveguides as near-field probes, evanescent-induced chiral fields, and chiroptical interaction in void slots. In all the cases we consider a metallic nanohelix as a chiral probe, though all the approaches can be extended to other kinds of chiral nanostructures as well as matter. Our results establish that chiral applications such as sensing and spectroscopy could be realized in standard integrated optics, in particular, with silicon-based technology.
\end{abstract}

\section{Introduction}

Chirality is the geometrical property by which an object cannot be superimposed on its mirror image. Even though this feature is ubiquitous in nature at the full range of scales, it is likely at the molecular level where it plays a key role from a practical point of view. Indeed, enantiomers, i.e., pairs of stereoisomers having opposite handedness or chirality, may have well different pharmacological effects \cite{FDA1992,Hutt1996,Smith2009}. Besides in the molecular structure \cite{Naaman2019}, chirality also applies to some elementary particles such as photons \cite{Lodahl2017}, electrons \cite{Gohler2011}, or phonons \cite{Zhu2018}. In the case of photons, and using classical notation, we shall refer to it as \textit{optical chirality}, which, just like the electromagnetic helicity \cite{Cameron2012,Alpeggiani2019}, has been recognized as a fundamental property of light characterizing the local handedness, or twistiness, of the electromagnetic field lines \cite{Tang2010,Bliokh2011}. Thus, circularly polarized light (CPL) is typically considered as the paradigmatic example of chiral light \cite{Tang2011}.

Information about the chirality of matter at the micro or even molecular scale can then be directly retrieved by making it interact with chiral light. There are a number of techniques to discriminate enantiomers, but the most commonly performed is the so-called \textit{circular dichroism spectroscopy} \cite{Berova}, which consists in determining the differential absorption rate of left (L) and right (R) CPL when propagating through a chiral medium \cite{Barron}. Likewise, the efficiency of this asymmetric chiroptical response is generally quantified in terms of the \textit{dissymmetry factor}:
\begin{equation}\label{gfactorAbsorption}
	g=2\left[\frac{A^+-A^-}{A^++A^-}\right],
\end{equation}
where $A^{\pm}$ is the absorption rate of L-CPL (+) and R-CPL (-). This establishes a standard metric to quantify the enantioselectivity and its possible enhancement with respect to CPL \cite{Hassey2006,Hendry2010}. Importantly, in the dipolar approximation, it has been proved that the dissymmetry factor is, in turn, proportional to the optical chirality density \cite{Tang2010,Tang2011}, thereby promoting an insightful pathway to control the chiroptical response via an all-optical approach, namely, regardless of the structural properties of the chiral analyte \cite{Rhee2013}. In this regard, it should be noticed that, although known for years, only recently it has been put forward general and closed expressions for both the optical chirality \cite{VazquezLozano2018}, as well for the electromagnetic helicity \cite{Alpeggiani2019}, of fields propagating through dispersive and even lossy media.

Interaction between chiral light and chiral molecules is in general extremely weak \cite{Barron}, owing to the large scale difference between the operational wavelength of the input light and the typical size of the molecules \cite{Rhee2013}. Nevertheless, since the original suggestion \cite{Tang2010} and preliminary demonstrations \cite{Tang2011,Hendry2010} of the existence of \textit{superchiral fields} (i.e., optical fields carrying chirality larger than that of CPL) to enhance the chiroptical interactions, the interest in this field has grown considerably. In particular, metallic nanostructures supporting plasmonic resonances at subwavelength dimensions have been proven to be well-suited platforms for strengthening the chiroptical effects \cite{Schaferling, Schaferling2012,Barr2018}, leading to local chiral enhancements by orders of magnitude \cite{Collins2017,Hentschel2017}. Further developments have led to improve and simplify the chiroptical schemes to achieve ultrasensitive enantiomeric detection, recognition, and separation \cite{Govorov2010}, for example, by stacking twisted planar metasurfaces \cite{Zhao2017}, or in arrays of both chiral \cite{Kang2017} and even achiral plasmonic nanostructures, such as spheres \cite{Kramer2017}, nanoslits \cite{Hendry2012}, or nanorods \cite{Meinzer2013,Nesterov2016}. More recently, high-index dielectric disks have also been suggested as suitable structures to boost near-field chirality \cite{Mohammadi2018}. Besides displaying lower absorption losses than its plasmonic counterparts, dielectric resonators are proving to be a promising platform with a number of important practical advantages for chiral sensing, spectroscopy, and enantioselectivity \cite{Pellegrini2018}: large areas of high and uniform-sign chirality (enantiopure enhancement) \cite{Solomon2019,Graf2019}, spectral accessibility and tunability of chiral interactions \cite{Mohammadi2019}, and switchability upon reversing the input polarization \cite{Zhao2019}.

In recent years there have been many efforts to implement sensing, spectrometry, and spectroscopy in photonic integrated platforms \cite{Estevez2012,Nie2017}. In all these cases, light is guided on a chip through lossless dielectric nanowires (i.e., waveguides with confinement dimensions of the order of $\lambda/2$, and propagation distances beyond the cm-scale), wherein some integrated elements are included to carry out the required optical processing. A similar conception may also be applicable to perform integrated chiral sensing and spectroscopy, since, up to now, all the approaches are considering arrangements in which the input light propagates through free space. While keeping the same working principle, the integrated approach should bring about numerous practical benefits, mainly if realized onto silicon-based platforms compatible with standard CMOS technology: low-cost and mass-volume production, repeatability, multiplexed detection, or integration with electronics, among others. Yet, in the context of integrated photonics, chiroptical effects have only been borne out as a consequence of the spin-orbit interaction of light for controlling near-field unidirectional excitation of guided modes \cite{Petersen2014,Coles2016,Gong2018}, leaving aside other interesting capabilities such as enantiomeric discrimination, chiral spectroscopy, or sensing. In this regard, it should be noted that, apart from the case of plane-wave propagating in free space, studies on optical chirality (and/or helicity \cite{Cameron2012}) in integrated platforms are scarce, with the notable exception of cylindrical dielectric waveguides, for which there are analytical solutions \cite{Picardi2018,Kien2017,Abujetas2018}.

In this work, we investigate the existence of enhanced optical chirality in lithographically-defined integrated dielectric waveguides, as well as its potential in chiral applications such as sensing, spectroscopy or filtering. In particular we focus on two of the most common guiding structures used in nanophotonics: strip and slot waveguides \cite{Saleh}. Even though the fundamental eigenmodes of these structures, typically referred to as transverse electric (TE) and transverse magnetic (TM), yield no chirality, we show that a net optical chirality of opposite signs can arise from the adequate superposition of two orthogonally polarized guided modes, analogously to the conventional case of L- and R-CPL. Building on this analysis, we show that such waveguides can be used as near-field tips and probes for excitation and read-out of chiral nanostructures, in arrangements equivalent to that routinely performed in circular dichroism measurements. More strikingly, we find a strong interaction between the evanescent field of the chiral guided modes and the chiral metallic sample, which is different depending on whether or not the optical chirality of the field matches the handedness of structure. These effects also allow us to disclose the active role of the absorption losses, recently introduced in the theoretical framework of optical chirality \cite{VazquezLozano2018}, as well as the relationship between the chirality and the transverse spin \cite{Bliokh2012,Alizadeh2015a}. Hence, besides suggesting the possibility of using photonic integrated platforms for chiral applications, these findings may lay the groundwork for a greater variety of chiroptical effects \cite{Nechayev2019,Petronijevic2019}.

\section{Results and discussion}

\subsection{Optical Chirality Density in All-Dielectric Integrated Photonic Waveguides}

We start by considering a guiding system consisting of a silicon nitride (Si$_3$N$_4$) waveguide with square cross-section that, for simplicity, is surrounded by vacuum. Silicon nitride is chosen as it is compatible with standard CMOS technology and enables guidance in an extremely broad wavelength regime \cite{Nie2017}, thus covering the relevant visible and near-infrared (NIR) spectral regions \cite{RomeroGarcia2013}. The waveguide can be suspended in air just by removing the substrate, as it is usually done in waveguide and cavity optomechanics. Moreover, by choosing a square cross-section we ensure that the two first guided modes (i.e., the TE and TM modes) appear to be degenerate. Unlike wires with circular-cross section \cite{Picardi2018,Kien2017,Abujetas2018}, there is no analytical solution for the propagating modes of square (or, in general, rectangular) dielectric waveguides, so we ought to calculate the corresponding guided modes by means of numerical simulations. This has been performed with the aid of the commercial 3D full-wave solver \textit{CST Microwave Studio} (see Methods section). After that, as done earlier when calculating the transverse spin of such guided modes \cite{EspinosaSoria2016a}, once determined the electric and magnetic components of each mode, we can straightforwardly obtain the optical chirality density from its general definition \cite{VazquezLozano2018}:
\begin{equation}\label{OC}
	C=\frac{\omega}{4c^2}\text{Im}{\left[\left(\varepsilon(\omega)\mu_{\rm eff}(\omega)+\varepsilon_{\rm eff}(\omega)\mu^*(\omega)\right){\bf E}\cdot{\bf H}^*\right]},
\end{equation}
wherein $c$ is the speed of light in free space, ${\bf E}$ and ${\bf H}$ are, respectively, the electric and magnetic complex field amplitudes, with the asterisk denoting complex conjugation, and $\varepsilon$ and $\mu$ are the standard permittivity and permeability of the medium. Additionally, to fully account for the dispersion of the guiding structure, we should use the real-valued effective material parameters $\varepsilon_{\rm eff}$ and $\mu_{\rm eff}$, depending on the angular frequency $\omega$ \cite{VazquezLozano2018}.

\begin{figure*}[t!]
	\centering
	\includegraphics[width=1\linewidth]{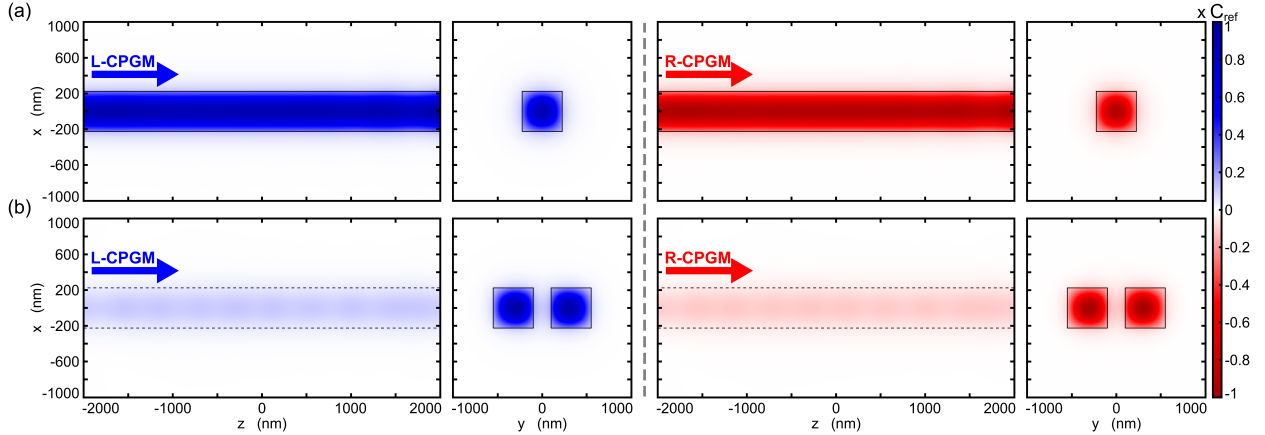}
	\caption{Cross-sectional maps of the normalized chirality density for L- and R-CPGM in silicon nitride (a) strip and (b) slot waveguides with square cross-section ($w=t=450$ nm) and $d_{\rm slot}=200$ nm, at NIR wavelengths. The normalization is carried out with respect to the maximum chirality, dubbed as $C_{\rm ref}$, for each configuration, and it will serve as a reference to quantify the corresponding chiral enhancement in the following results.}
	\label{fig_01}
\end{figure*}

Figure \hyperref[fig_01]{1a} shows the optical chirality density for right- and left-handed circularly polarized guided modes (CPGMs) in a silicon nitride strip waveguide, with equal width and thickness, $w=t=450$ nm, at NIR frequencies. Notice that, in principle, lower wavelengths (i.e., those ranging from visible to the near-UV spectral region) would be more suitable for chiral spectroscopy. Nonetheless, the transverse profile of the modes will be similar at any wavelength below the cut-off, so these results would qualitatively apply to the whole wavelength range supporting the TE and TM guided modes. Just like CPL propagating in free space, CPGMs may be built by superposing both degenerate TE and TM eigenmodes with a phase shift of $\pi/2$ between them. In this way, they result in a locally nonzero optical chirality within the waveguide core, being positive or negative depending on whether the CPGM is left- or right-handed, respectively. This is a feature that could be expected by considering that the waveguide core squeezes propagating light in a thin cross-section of the order of $(\lambda/2n_{\rm eff})^2$, being $n_{\rm eff}$ the effective refractive index of the corresponding guided mode. Noticeably, the evanescent tail of the propagating CPGM through the vacuum region surrounding the waveguide still yields chirality as well. This allows us to suggest the possibility of performing chiral spectroscopy or sensing just by placing chiral objects in such a near-field region. As evident, the local chirality of the evanescent field is much lower than in the core, but still, its interaction with chiral structures can be even stronger than that of normal incidence, as will be shown in the next section. Moreover, since all-dielectric waveguides are inherently lossless, the optical chirality conveyed by the CPGMs can be transported over very long distances, thereby enlarging the active region where realizing chiral light-matter interaction.

\begin{figure}[t!]
	\centering
	\includegraphics[width=0.55\linewidth]{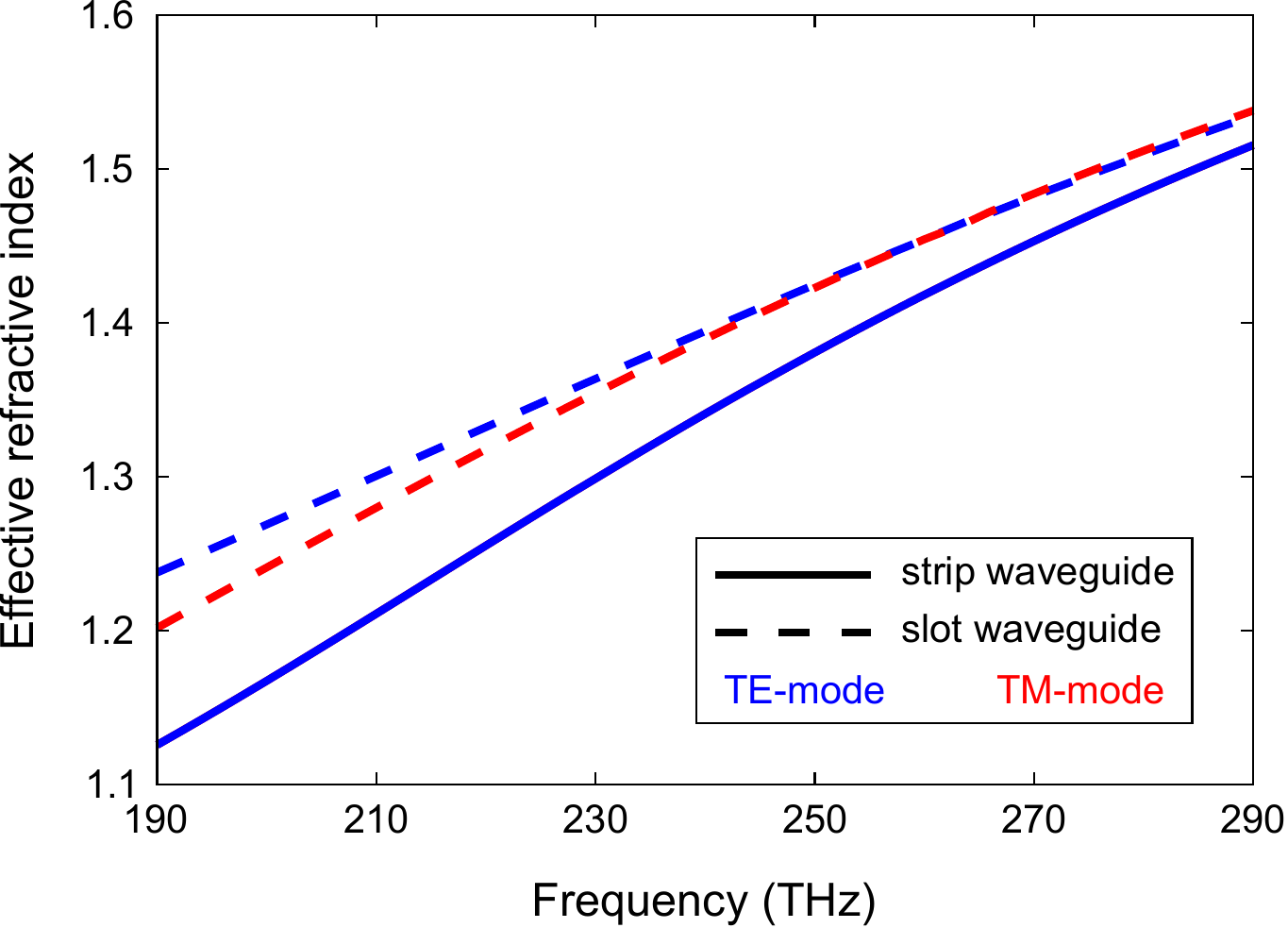}
	\caption{Effective refractive index of the fundamental TE and TM eigenmodes of a silicon nitride strip (solid line) and slot (dashed line) waveguides with square cross-section ($w=t=450$ nm) and $d_{\rm slot}=200$ nm. In both cases the surrounding medium is assumed to be vacuum.}
	\label{fig_02}
\end{figure}

Alternatively, dielectric slot waveguides are practical structures that enable guiding and strong field confinement in a thin low-index region separating two strip waveguides \cite{Almeida2004}. This makes them useful for biosensing \cite{Barrios2007}, and thus amenable for the experimental realization of chiroptical applications. However, in contrast to the previous case, here the slot breaks the symmetry of the system so it is not possible to get an exact degeneracy between the TE and TM modes. It is though feasible to design the slot waveguide in such a way that the TE and TM modes have the same effective index for (or up to) a certain wavelength, namely, ensuring a sufficient degree of coherence for the CPGMs so that they can propagate along the required operational distance. As shown in Figure \hyperref[fig_02]{2}, for a square cross-sectional silicon nitride waveguide with $w=t=450$ nm, and $d_{\rm slot}=200$ nm, this overall degree of coherence between TE and TM modes is attained when $f\approx250$ THz ($\lambda\approx1200$ nm). With this restriction in mind, we performed similar calculations as in the former case, whose results are depicted in Figure \hyperref[fig_01]{1b}. It can be seen that both L- and R-CPGM yield a roughly uniform distribution of optical chirality in the slot region, which should be extremely useful for chiral applications.

The optical chirality maps shown in Figure \hyperref[fig_01]{1} will be used as the reference for setting the corresponding enhancement when the chiral sample is included. In this regard, it is worth pointing out that, unlike in the case of plane-wave propagation in free space, wherein superchirality is well defined with respect to the chirality of CPL, in the case of waveguiding systems the situation can be slightly problematic as it would depend on the specific structure. So, in lieu of superchirality, we shall refer to it simply as the enhanced chirality, whose enhancement factor is relative to the reference chirality $C_{\rm ref}$, i.e., the maximum chirality for each specific configuration.

Now that we have shown that planar dielectric waveguides may support chiral light propagation, we will look into three different scenarios enabling integrated chiral applications: dielectric waveguides as near-field probes for excitation and read-out of chiral light-matter interaction (\textit{in-gap configuration}); near-field optical chirality induced by evanescent fields neighboring waveguiding structures (\textit{on-top configuration}); and chiroptical interaction in slot waveguides (\textit{slotted configuration}). For simplicity, and without any loss of generality, throughout this work we are going to consider a metallic nanohelix as a sample structure since it is often regarded as the prototypical example of chiral object \cite{Gansel2009,Schaferling2014,Esposito2015,Ji2016,Wozniak2018,Hoflich2019}. It will lead to a differential absorption (or, equivalently, transmission) of L- and R-CPGM, depending on the handedness of the metallic nanohelix. For comparison purposes, we will analyze the resonant behavior of this structure modeled either as a perfect electric conductor (PEC) or silver, whose material parameters are taken from Ref. \blue{60}. At any rate, we stress out that any other kind of chiral structure or material could be used as well.

\subsection{Probing the Chiroptical Response in Dielectric Strip Waveguides: Normal Incidence (in-gap configuration) and Evanescent-Induced Chiral Interaction (on-top configuration)}

\begin{figure}[t!]
	\centering
	\includegraphics[width=0.55\linewidth]{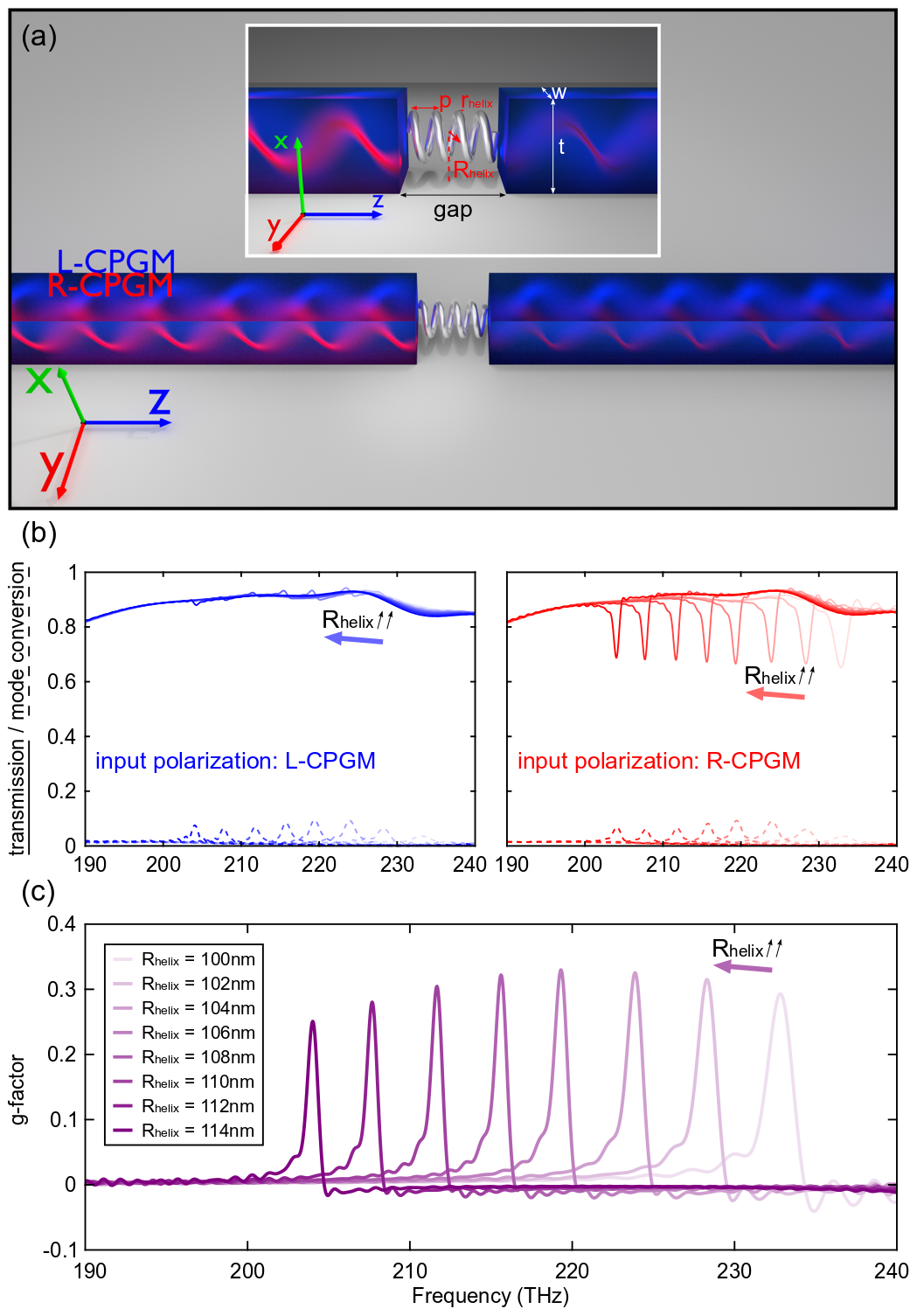}
	\caption{(a) Scheme of the arrangement for chiral sensing and spectroscopy enabled by a dielectric strip waveguide as a local probe for excitation and read-out under normal incidence with a gap width set to $240$ nm (\textit{in-gap configuration}). The chiral metallic structure is modeled as a right-handed $4$-turns single helix made of PEC, with geometrical parameters fixed to $p=50$ nm, $r_{\rm helix}=20$ nm, and $R_{\rm helix}$ is varied from $100$ to $114$ nm. (b) Numerical simulation of transmission and mode conversion spectra for different values of $R_{\rm helix}$. (c) Dissymmetry factor.}
	\label{fig_03}
\end{figure}

First we investigate a chiral metallic scatterer embedded in a narrow gap separating two silicon nitride waveguides with square cross-section ($w=t=450$ nm). The chiral nanostructure consists of a $4$-turns single helix placed in between the waveguides, being centered and orientated along the propagation axis in the $z$-direction (see Figure \hyperref[fig_03]{3a}). As aforementioned, in this first approach, the metal is simply modeled as a PEC. This would give us some hints to understand the chiral behavior of the helix on account of its geometry itself, thereby dismissing any potential influence of dispersion or absorption effects. Regardless of the number of turns, a full geometrical description of this complex object is determined by the following set of structural parameters: helix radius ($R_{\rm helix}$), pitch ($p$), and helix wire radius ($r_{\rm helix}$), apart from the handedness, given by a relative phase shift between the $x$ and $y$ coordinates. By properly engineering each of these parameters one can gain control over the main spectral features, tuning the resonance position and depth, as well as the optical activity of the metallic chiral structure (see Refs. \blue{58} and \blue{59} for further details). In this case, we shall consider a right-handed helix with geometrical parameters fixed to $p=50$ nm, $r_{\rm helix}=20$ nm, and $R_{\rm helix}$ set as the control parameter, varying from $100$ to $114$ nm. Notice that this choice of the variable parameter is done so that we can keep constant the gap width, in this case to $240$ nm.

As mentioned above, we can excite either L- and R-CPGM in the input waveguide by controlling the relative phase shift between the TE and TM modes (see Methods section). Then, using an end-fire coupling setup, the light exiting the input waveguide will excite the chiral structure in the near-field, here acting as a probe. The output waveguide will collect the light both scattered by the structure and transmitted from waveguide to waveguide \cite{EspinosaSoria2016b,EspinosaSoria2018}, thus enabling the read-out of the chiroptical interaction. To account for the chiral spectral response, we first represent the transmission and mode conversion spectra for L- and R-CPGM (see Figure \hyperref[fig_03]{3b}). From this, it can be seen that, while L-CPGM is nearly transparent to the presence of the helix, R-CPGM exhibit a narrow transmission dip depending on $R_{\rm helix}$, thereby providing clear evidence of the underlying optical activity brought about when the handedness of the CPGM coincides with that of the chiral structure. These signals directly translate into an asymmetric response on the transmission spectra, whose strength is generally quantified by the chiral dissymmetry factor $g$ (see Figure \hyperref[fig_03]{3c}). In this regard it should be noted that, along with the definition given in Eq. \eqref{gfactorAbsorption}, involving the differential absorption rate, the dissymmetry factor may also be equivalently characterized in terms of the differential transmission. Thus, concerning our proposed scheme, such a definition will allow us from now on for a more direct and intuitive description of the chiroptical interaction. It can be seen that the peak values of $g$ ($\approx 0.3$) are attained at the helix resonance, which does depend on the radius $R_{\rm helix}$. This then confirms the chiral response of the system in a region with a foot-print $<1$ $\mu$m$^2$. Remarkably, as shown in Figure \hyperref[fig_03]{3b}, the strong chiral interaction at resonance also results in a small, but nonzero, mode conversion between L- and R-CPGM, and vice versa.

\begin{figure}[t!]
	\centering
	\includegraphics[width=0.55\linewidth]{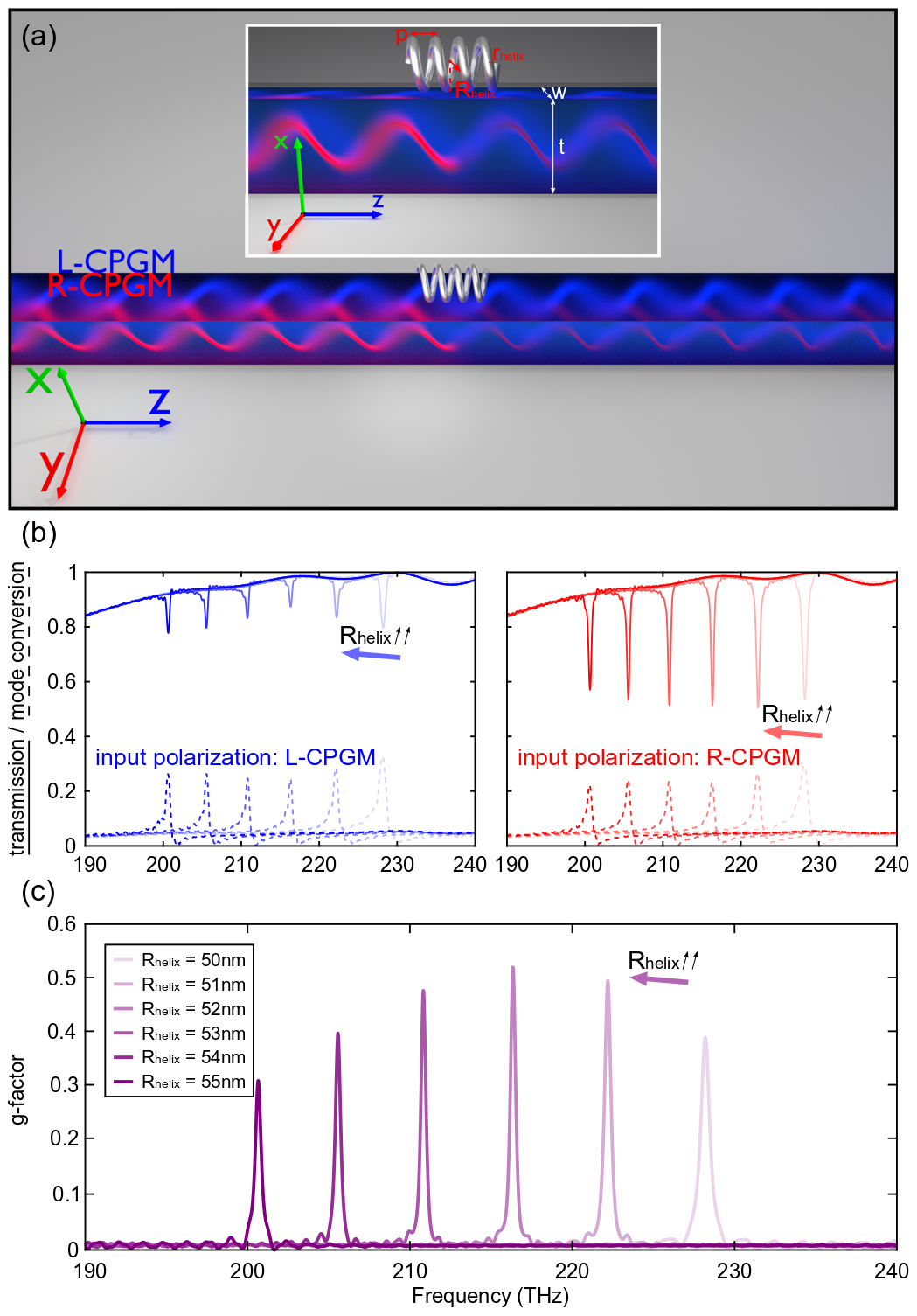}
	\caption{(a) Scheme of the arrangement for the evanescent-induced chiral response in a dielectric strip waveguide (\textit{on-top configuration}). The characterization of the chiral metallic nanohelix is the same as in Figure \hyperref[fig_03]{3}, but with $R_{\rm helix}$ ranging from $50$ to $55$ nm. (b) Numerical simulation of transmission and modal conversion spectra for different values of $R_{\rm helix}$. (c) Calculated dissymmetry factor.}
	\label{fig_04}
\end{figure}

A similar analysis holds for the near-field chirality induced by the evanescent field neighboring the silicon nitride waveguide over which a metallic chiral nanostructure is placed. The scheme of this approach is depicted in Figure \hyperref[fig_04]{4a}. Again, maintaining the same geometrical parameters for the nanohelix and modeling it as a PEC, we obtain, via numerical simulations, the spectral behavior of the transmission of L- and R-CPGM for different radii of the helix, now ranging from $50$ to $55$ nm (see Figure \hyperref[fig_04]{4b}). From these results, and taking into account the above considerations, we can directly calculate the dissymmetry factor $g$ (see Figure \hyperref[fig_04]{4c}). It can be seen that the peaks of $g$ (or, analogously, the dips in transmission) become narrower and higher in comparison with the previous case. This is quite striking since the fields in the evanescent region are much weaker than in the gap spacing two waveguides, wherein they are essentially similar to those in the waveguide core \cite{EspinosaSoria2016b}. Yet, this configuration still retain common features with respect to the previous case, in particular, the peak position (or resonance frequency) dependent on the helix radius, and the intermodal conversion between L- and R-CPGM, and vice versa, as a result of the interaction between the CPGMs and the chiral structure (Figure \hyperref[fig_04]{4b}).

\begin{figure*}[t!]
	\centering
	\includegraphics[width=1\linewidth]{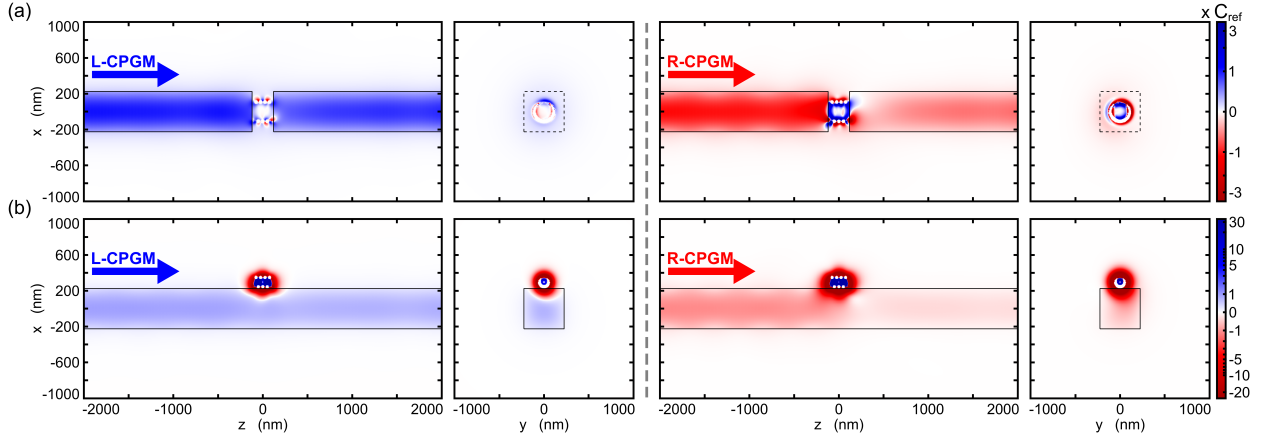}
	\caption{Cross-sectional maps of the optical chirality density for L- and R-CPGM in silicon nitride strip waveguides for (a) the in-gap configuration (Figure \hyperref[fig_03]{3a}) and (b) the on-top configuration (Figure \hyperref[fig_04]{4a}). The chiral metallic structure is modeled as a PEC, and the geometrical parameters are the same as in Figures \hyperref[fig_03]{3} and \hyperref[fig_04]{4}, setting (a) $R_{\rm helix}=108$ nm, and (b) $R_{\rm helix}=52$ nm. In both cases we consider the working frequency $f=f_{\rm res}\approx 216$ THz. The values are normalized with respect to $C_{\rm ref}$, i.e., the maximum chirality attainable for each configuration when ignoring the presence of the chiral structure.}
	\label{fig_05}
\end{figure*}

For a more intuitive description, we plot the cross-sectional map of optical chirality density from the corresponding electromagnetic field distributions (see Figure \hyperref[fig_05]{5}), which, in turn, are obtained via numerical simulations (see Methods section). For practical reasons, we have to decide a specific set of parameters to perform the graphical representation, including the working frequency. By looking at the results shown in Figures \hyperref[fig_03]{3c} and \hyperref[fig_04]{4c}, the best trade-off for a fair comparison is such that $R_{\rm helix}=108$ nm for the in-gap configuration, and $R_{\rm helix}=52$ nm for the on-top configuration, since in both cases there exists a well-localized resonance at approximately $216$ THz. It should be noticed that this map brings about complementary information to be accounted for. In particular, it provides a clear visualization of the resonant behavior of the nanohelix. Indeed, as shown in Figure \hyperref[fig_05]{5}, most of the optical chirality density is tightly localized and strongly enhanced in the surroundings of the chiral structure, reaching a value of up to $30$ times $C_{\rm ref}$, i.e., the maximum chirality attainable when the chiral particle is ignored (see Figure \hyperref[fig_01]{1}). Remarkably, both the enhancement and the localization are even larger when the input mode possesses the same handedness as the chiral nanohelix, namely, for R-CPGM. This translates into a higher absorption by the particle (or a lower transmission at the output port), thus exhibiting a true chiroptical effect. Furthermore, regardless of the input mode, the sign of chirality remains uniform in both the outer and the inner region of nanohelix. In this regard, it is worth observing that, inside the helix, the chiral sign is always opposed to that of its handedness. This fact becomes even more evident for the on-top configuration, in which the local optical chirality gets strikingly higher than in the in-gap configuration.

\subsection{Further Considerations for a Realistic Approach of Chiroptical Applications in Integrated Platforms}

\subsubsection{Beyond the Ideal PEC Model: Drude-Lorentz Materials}

\begin{figure*}[t!]
	\centering
	\includegraphics[width=1\linewidth]{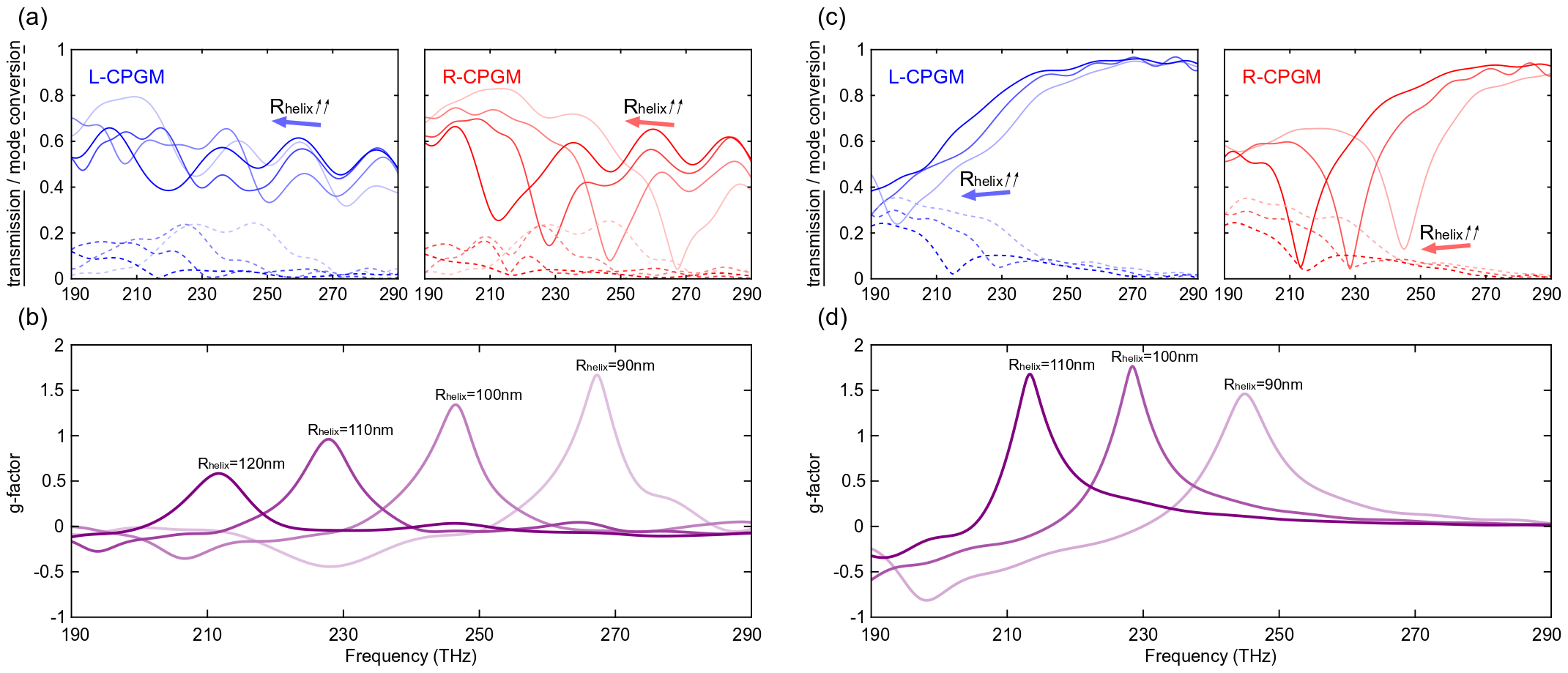}
	\caption{(a) Numerical results of transmission and mode conversion spectra and (b) calculated dissymmetry factor for different values of $R_{\rm helix}$ for the in-gap configuration. Panels (c) and (d) plot the same for the on-top configuration. Note that in all panels we are considering a right-handed silver nanohelix with $p=70$ nm and $r_{\rm helix}=10$ nm. For the in-gap configuration the gap width is fixed to $300$ nm and the helix has $4$ turns, whereas for the on-top case it has $16$ turns.}
	\label{fig_06}
\end{figure*}

\begin{figure*}[t!]
	\centering
	\includegraphics[width=1\linewidth]{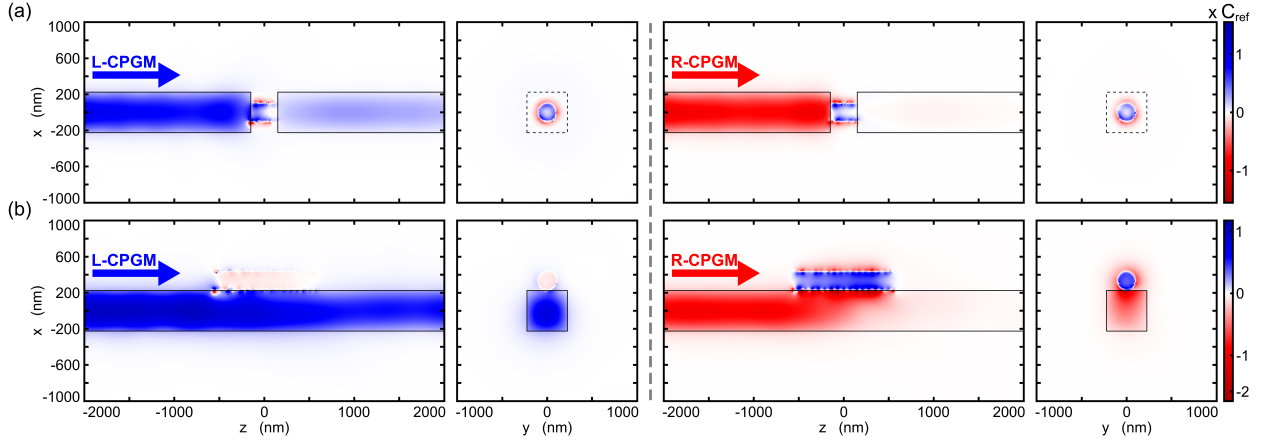}
	\caption{Cross-sectional maps of the optical chirality density for L- and R-CPGM in silicon nitride strip waveguides for (a) the in-gap configuration (Figure \hyperref[fig_03]{3a}) and (b) the on-top configuration (Figure \hyperref[fig_04]{4a}). The chiral metallic structure is modeled as a right-handed silver nanohelix with geometrical parameters as indicated in Figure \hyperref[fig_06]{6}. In both cases the working frequencies are fixed to be that of resonance for $R_{\rm helix}=100$ nm, i.e., (a) $f=f_{\rm res}\approx 247$ Thz, and (b) $f=f_{\rm res}\approx 228$ THz. The values are normalized with respect to $C_{\rm ref}$.}
	\label{fig_07}
\end{figure*}

For the sake of simplicity, thus far we have assumed the chiral probe particle as made of a PEC. However, this will not be the case in a real system, where at optical frequencies metals display a Drude-Lorentz-like response in which dispersion and absorption losses should be included. To account for this, we have further repeated our simulations considering a silver helix with material parameters taken from Ref. \blue{60}. Numerical results are summarized in Figure \hyperref[fig_06]{6}. In order to compare with the case above, for the in-gap configuration, we have considered again a right-handed $4$-turns single helix. Nonetheless, for a better representation of the results, small modifications in the geometrical parameters have been carried out, setting the gap width to $300$ nm, $p=70$ nm, $r_{\rm helix}=10$ nm, with $R_{\rm helix}$ varying from $90$ to $120$ nm. On the other side, for the on-top configuration we have used a helix with $16$ turns, so that we can leverage further the chiroptical interaction induced by the evanescent tail of the CPGMs. Regardless of structural specifications of the chiral scatterer, in both cases, the system still retains the same general features mentioned above, i.e., chiral-like resonances whose frequency depends on the helix radius, and the nonzero mode conversion between L- and R-CPGM, and vice versa. However, likely due to the absorption losses, the transmission dips, and consequently, the peaks in the dissymmetry factor spectra become wider. Moreover, $g$ also exhibits opposite chiral behavior at frequencies just before the resonant peak. Despite that, it is worth emphasizing that the chiroptical responses characterized by the dissymmetry factor are greatly increased in comparison with the previous results for PEC made helices. Once again, we illustrate this fact by means of the chirality density cross-sectional map shown in Figure \hyperref[fig_07]{7}. Even though the enhancement is obviously lower than for the case of PEC helices, the contrast ratio between L- and R-CPGM at the output waveguide is now significantly higher. Namely, whilst the chirality density is roughly uniformly preserved along the entire waveguide for L-CPGM, the excitation of the chiral structure by the R-CPGM is so strong that the chirality is completely absorbed at the output port, a fact that can be attributed to the active role played by the absorption losses \cite{VazquezLozano2018}.


\subsubsection{Enhancing the Evanescent Effect: Slotted Configuration}

\begin{figure}[t!]
	\centering
	\includegraphics[width=0.55\linewidth]{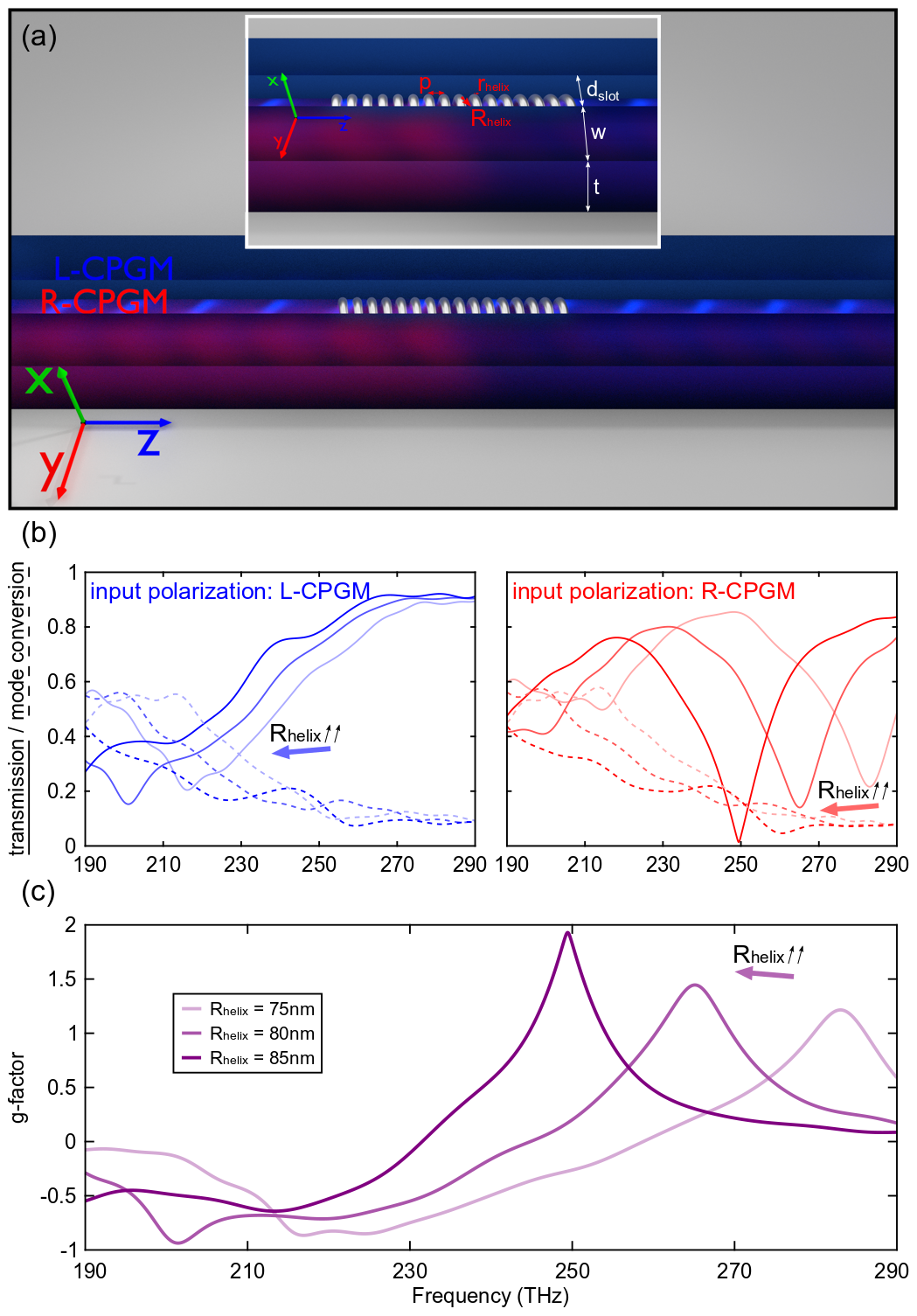}
	\caption{(a) Scheme of the arrangement for chiral sensing and spectroscopy enabled by a dielectric slotted waveguide with a slot width of $d_{\rm slot}=200$ nm (\textit{slotted configuration}). The chiral metallic structure is modeled as a right-handed $16$-turns single nanohelix made of silver, with geometrical parameters fixed to $p=70$ nm, $r_{\rm helix}=10$ nm, and $R_{\rm helix}$ ranging from $75$ to $85$ nm. (b) Numerical simulation of transmission and mode conversion spectra for different values of $R_{\rm helix}$. (c) Dissymmetry factor.}
	\label{fig_08}
\end{figure}

Slot waveguides, consisting of a thin air channel placed between two dielectric waveguides \cite{Almeida2004}, present important advantages in a multitude of integrated photonics applications, in particular, in optical biosensing and spectroscopy \cite{Estevez2012,Barrios2007}. Motivated by this, we also analyze chiroptical light-matter interactions in these kind of configurations. Indeed, the slotted approach can be regarded as an improved version of the on-top configuration in which we place a second waveguide close to the first one in order to produce more intense fields in the void region close to the waveguide interface. Therefore, as in the case of the on-top configuration, the main benefit of the slotted configuration relies on the long active region in which the chiroptical interaction takes place. 

Based on this, we consider two square cross-sectional silicon nitride waveguides separated by a slot of width $d_{\rm slot}=200$ nm, where the optical field is strongly confined. Here we place the chiral analyte, in this case a $16$-turns silver nanohelix (see Figure \hyperref[fig_08]{8a}). The rest of geometrical parameters are the same that previously used in Figure \hyperref[fig_06]{6}. From the numerical results shown in Figure \hyperref[fig_08]{8} we observe that, like in the cases above, the transmission spectra display a sharp dip at the resonance frequency for the CPGM that matches the chirality of the helix. This translates into a peak in the corresponding dissymmetry factor spectrum, which, owing to the absorption losses of the material, has a broad spectral width.

A key issue arising with the slotted configuration is that the system does not display a $C_4$ rotational symmetry around the waveguide axis, which prevents TE and TM modes to be degenerate. This means that the chirality of the CPGMs will change when propagating through the slotted waveguide. Still, in the simplest case in which the waveguide is surrounded by a homogeneous low-index medium, e.g., vacuum, the slot waveguide can be designed to be fully degenerate for (or up to) a given wavelength (see Figure \hyperref[fig_02]{2}) so that purely L- or R-CPGM can be generated and propagated for realizing chiral interaction. The $C_4$ symmetry also breaks when a substrate is introduced, and, as discussed below, this poses some limits in practice.

\subsubsection{Breaking the $C_4$ Rotational Symmetry: Polarization Beat Length and Its Effect on Chiroptical Applications}

All-dielectric waveguides with square cross-section completely surrounded by vacuum support TE and TM modes fully degenerate, meaning that the chirality conveyed by the CPGMs will be maintained no matter how long the waveguide is. Notice that this kind of suspended dielectric nanowires can be technologically realized by releasing the waveguide core using standard processes, for instance, in the case of silicon-based materials, by employing an HF bath that removes the silica substrate as usually done in cavity and waveguide optomechanics. However, this limits the maximum distance for which the chirality is maintained, as very long waveguides may eventually break down and collapse. This can be solved by using a substrate, so that there is no limit on the maximum guiding length, as long as we ensure low propagation losses, and providing a support for the addition of chiral scatterers (as the helix used in this work) or even chiral chemical compounds.

\begin{figure}[t!]
	\centering
	\includegraphics[width=0.55\linewidth]{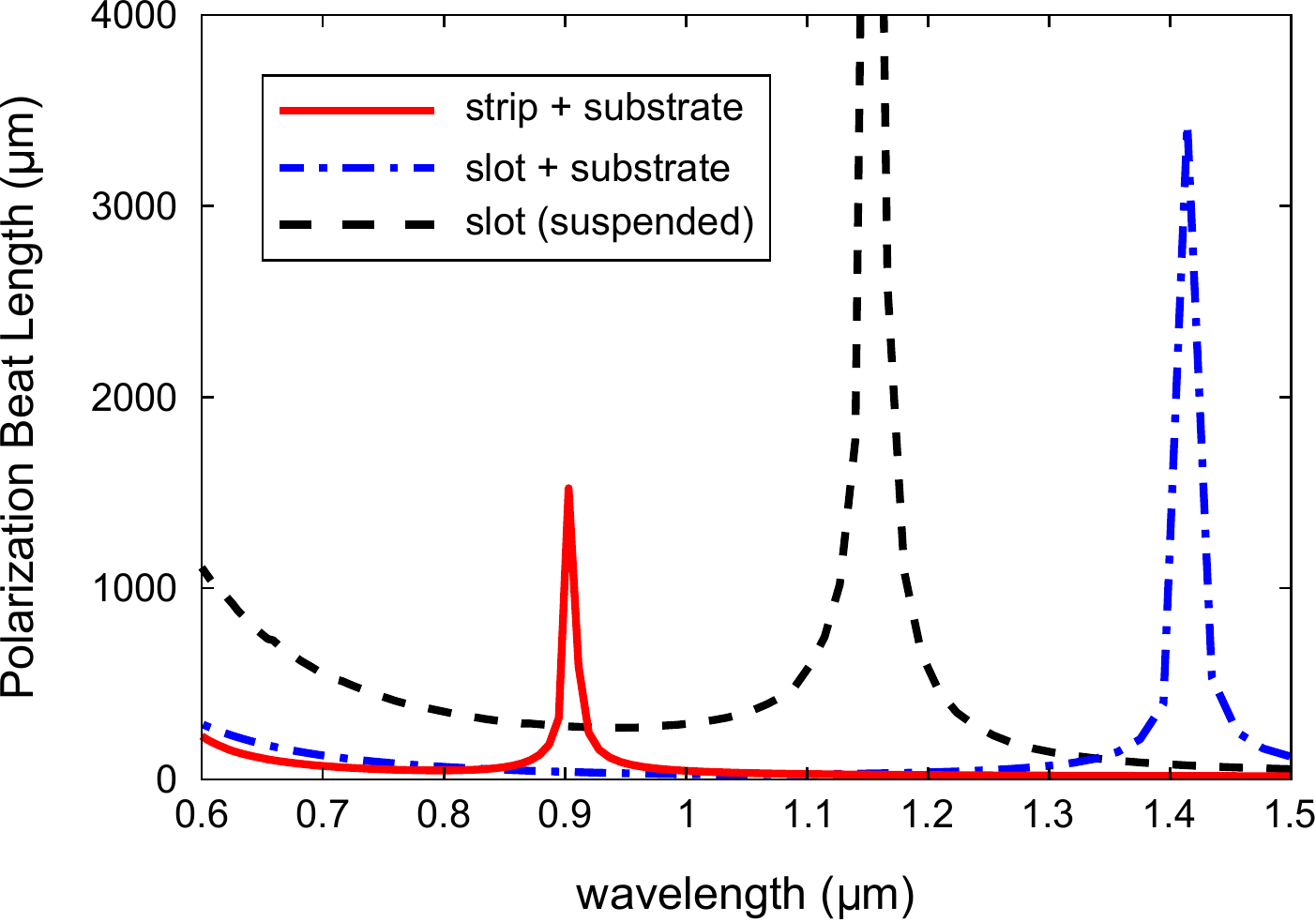}
	\caption{Polarization beat length (${\rm PBL}$) of a silicon nitride strip (solid line) and slotted (dashed lines) waveguide as a function of the wavelength. For the strip case, the rectangular cross-sectional waveguide ($w=350$ nm and $t=300$ nm) is placed on a silica substrate. In the slotted case, for comparison, the ${\rm PBL}$ is plotted with and without substrate.}
	\label{fig_09}
\end{figure}

When the substrate is added, the $C_4$ rotational symmetry of the system is broken, and the TE and TM modes are no longer fully degenerate. In such a case, it is worth assessing how long the sign of the chirality can be kept in our system, thereby determining the maximum active length to carry out the required chiral processing. To characterize the degree of degeneracy between the TE and TM eigenmodes, and its influence on the coherent propagation through the waveguide structure to generate the required chiral CPGMs, we use the so-called \textit{polarization beat length}:
\begin{equation}\label{PBL_definition}
	{\rm PBL}=\frac{\lambda}{n_{\rm TE}-n_{\rm TM}}.
\end{equation}
This is a parameter typically used to check the birefringence of optical waveguides \cite{Filippov1990}, accounting for the fabrication deviations from the nominal values of the waveguide as well. Specifically, it fixes the distance traveled by the corresponding guided mode after which the original polarization state is recovered. Namely, this means that the chirality sign can be kept uniform over a distance of ${\rm PBL}/2$. Therefore, in the cases of a suspended slotted waveguide or a dielectric wire with substrate, one can work out a design so that the TE and TM modes are fully degenerate at a certain frequency (for which ${\rm PBL}\to\infty$), and look into the bandwidth around that frequency so that it achieves a ${\rm PBL}$ length over a given threshold. In Figure \hyperref[fig_09]{9} we plot the PBL for both strip (solid line) and slot (dashed line) waveguides. To do that, we first calculated the effective refractive index for the TE and TM modes in each case by means of a 3D full-wave solver (see Methods section). In the strip case we consider a realistic system: a conventional silicon nitride wire with rectangular cross-section, being $t=300$ nm (available in commercial wafers) and $w=350$ nm, on top of a silica substrate. On the other side, for the slotted case we choose square waveguides with the same dimensions as that indicated in Figure \hyperref[fig_02]{2}, and, for comparison, we account for the cases with and without substrate. In the former scheme, we find that the TE and TM modes have the same index at a wavelength of $900$ nm, with ${\rm PBL}\approx 1.5$ mm over a bandwidth of nearly $50$ nm. Of course, this result is far from being satisfactory for a real application and it should be further optimized, by changing the dimensions as well as the characteristics of the cladding. Nevertheless, it serves to show the feasibility of such an approach taking into account the substrate. In fact, from a simple comparison with the slot waveguide, including the silica substrate, we observe a much better performance, increasing the ${\rm PBL}$ peak magnitude to $3.5$ mm, located in this case at $1400$ nm, and slightly widening the bandwidth. Finally, in the ideal case removing the substrate we can further enhance both the magnitude and the bandwidth of the ${\rm PBL}$ peak, centered at $1150$ nm. Interestingly, all the ${\rm PBL}$ peaks appear located in a spectral range that, as reported in the recent literature, is highly appropriate for realizing chiroptical spectroscopy, especially when plasmonic nanostructures are involved (see, e.g., examples shown in Ref. \blue{19}). Therefore, we have shown that, even with a substrate, realistic dielectric waveguides may support chiral modes in a significant wavelength range and over large propagation lengths, enabling chiral applications in active regions able to reach up to the mm-scale.

\section{Conclusions}

In summary, we have examined three possible configurations to get chiroptical responses enabled by all-dielectric strip and slotted waveguides in integrated photonic systems. To this aim, we have taken under consideration a metallic nanohelix acting as a chiral sample (a sort of metamolecule). For simplicity, the metal has been first modeled as a PEC, so that the chiral interaction is only due to the geometrical aspects, dismissing additional dispersive or absorption effects. Numerical results show that, in each of the configurations, there exists a very well localized single chiral-like resonance in the transmission spectra depending on the helix radius, which is chosen as the control parameter for sweeping. Consequently, the dissymmetry factor exhibits a narrow peak at such frequencies. Furthermore, intermodal conversion is also present everywhere. These features are illustrated by means of the chirality cross-sectional map, where it can be observed the strong enhancement in the chiral excitation of the nanohelix when its chirality coincides with that of the input CPGM.

For the in-gap configuration the system becomes ultra-compact, with an active area wherein realizing the chiroptical interaction smaller than $1$ $\mu$m$^2$. In contrast, for both the on-top and the slotted configurations, the interaction lengths can be made much larger, which may be highly useful to sort the chirality of chemical and biological compounds in the context of lab-on-a-chip technologies. In these cases, breaking the rotational symmetry of the system, either by adding a substrate or in the slotted waveguides themselves, would result in TE and TM modes no longer fully degenerate, limiting the length over which a certain sign of chirality can be kept uniform. Still, we have argued that, under a proper design of the waveguide and the cladding characteristics, the sign of the chirality could be conserved up to mm-scale distances over a wide bandwidth.

Hence, our results suggest that chiral applications such as sensing and spectroscopy, as well as polarization filtering, would be feasible using integrated photonic waveguides. Specifically, using silicon-compatible materials and technology would enable low-cost and large-volume production of chips for massive parallel detection, recognition, and separation of chiral molecules, thus completing current approaches for biosensing or spectroscopy in integrated platforms. Notice that far from providing fully optimized geometries, we only sought to show the viability of this working principle enabled by integrated photonic waveguides. Nonetheless, by comparing our results, in particular those of silver made helices, with those found in the existing literature (see, e.g., the overview table shown in the supplementary material of Ref. \blue{59}), it can be seen that the dissymmetry factors obtained here are comparable and even better than those already reported for freely propagating optical fields.

Although throughout this work we only focused on the chiroptical behavior of a metallic nanohelix, in real applications this would be extensible to other geometries and materials more relevant, specifically to chiral molecules. Likewise, taking into account earlier schemes, plasmonic nanostructures might be added in order to enhance the chiroptical response when a certain substance is applied \cite{Zhang2019}, thus leading to hybrid chiral plasmonic-photonic circuits. Somehow connected with this, it is worth noticing that the chiral structure considered here has only been used as a chiral sample to probe the chiroptical interactions. However, in view of the chirality maps shown above, it can also be envisaged as a near-field chiral enhancer, retaining the most important features for practical purposes \cite{Pellegrini2018}, namely, a large area of high and uniform-sign chirality, spectral accessibility, tunability, and switchability. In fact, we observe a clear analogy between the near-field chirality induced by the evanescent field and the scheme proposed in Ref. \blue{55}, in which the optical chirality is longitudinally confined within the helical structure.

Following the current trends, the next steps should be aimed at simplifying both the chiroptical measurement and enhancement schemes. In this regard, apart from investigating other designs of chiral structures simpler to fabricate \cite{Schaferling2016}, we also suggest exploring the recently introduced full-reconfigurable silicon-based on-chip wireless photonic systems \cite{Meca2017}, as a promising alternative platform for realizing chiral applications, such as sensing, spectroscopy, and enantioselectivity, from a lab-on-a-chip perspective.

\section{Methods}

All the numerical simulations and calculations have been performed with the aid of the commercial software packages CST Microwave Studio and MATLAB. In particular, for the simulations, we used the 3D full-wave solver to implement the finite-integration technique in the time domain. The whole model (waveguides and nanohelix) was meshed using a hexahedral grid with $16$ cells per wavelength, except near the metallic helix where additional refinement was necessary to properly resolve the small inclusions. Open boundary conditions (perfectly matched layers) have been chosen for all external facets, noticing that the background medium is the vacuum. The input excitation was provided by means of standard waveguide ports. In order to obtain the transmission as well as the mode conversion spectra at the output port, we sequentially stimulated both the TE and TM modes, recording their corresponding S-parameters. Then, applying the superposition principle, we translated these signals to the circular polarization basis, allowing the calculation of the corresponding dissymmetry factor for each case. On the other side, the dispersion relation of the fundamental TE and TM guided modes was obtained by simulating the guided modes only at the ports, for each central frequency. This calculation provides useful information about several waveguide port characteristics, in particular, the effective dielectric constant.

\begin{acknowledgement}
The authors thank S. Lechago for valuable comments and technical support with the numerical simulations.
This work was partially supported by funding from the European Comission Project THOR H2020-EU-829067. A.M. also acknowledges funding from Generalitat Valenciana (Grant No. PROMETEO/2019/123) and Spanish Ministry of Science, Innovation and Universities (Grant No. PRX18/00126).
\end{acknowledgement}

\section{Graphical TOC Entry}
\begin{center}
	\fbox{\parbox[][147pt][c]{260pt}{\centering\includegraphics[width=1\linewidth]{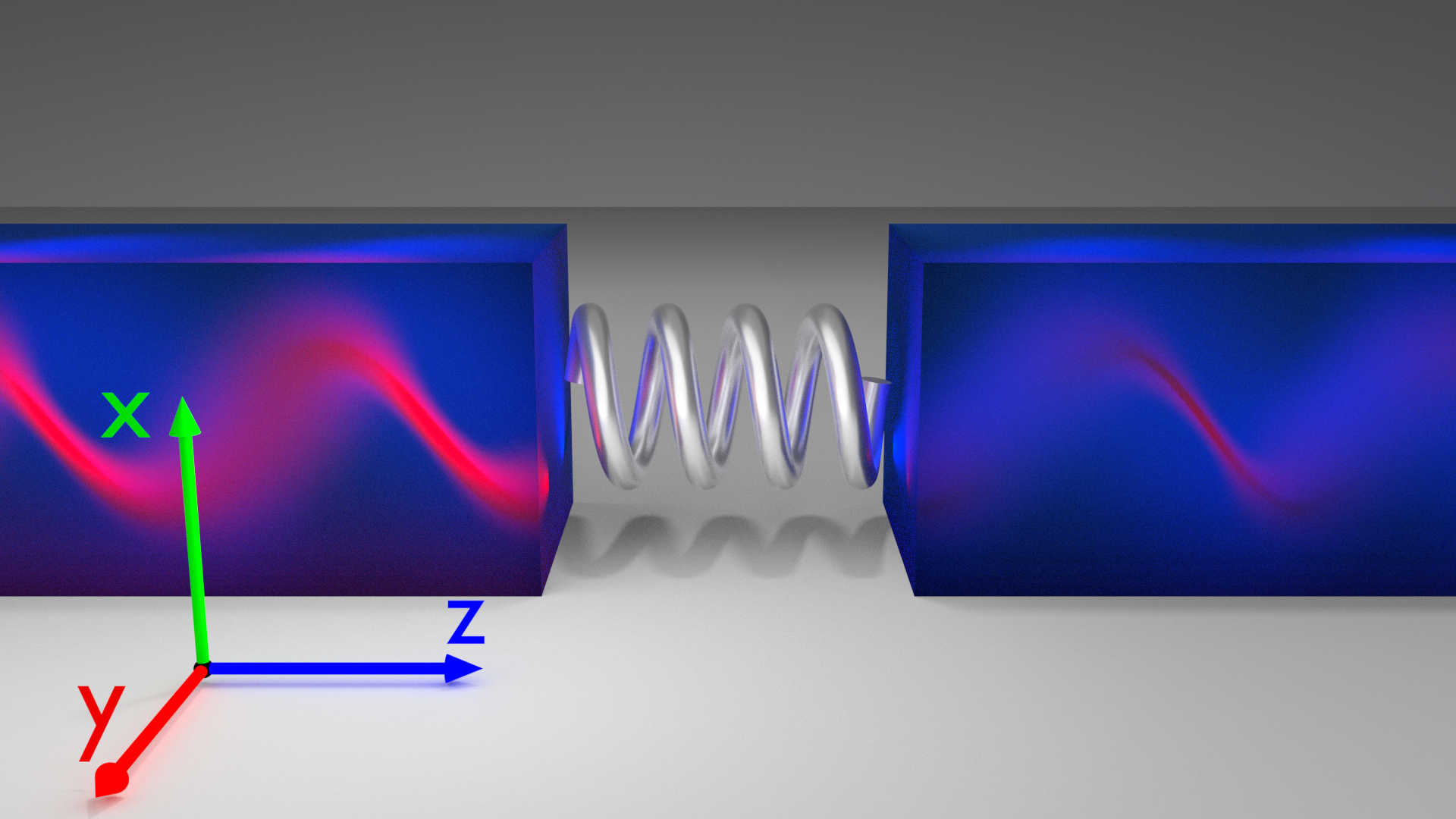}}}
\end{center}

\begin{thebibliography}{100}
	
	\bibitem{FDA1992} US Food and Drug Administration. FDA'S policy statement for the development of new stereoisomeric drugs. \href{https://doi.org/10.1002/chir.530040513}{\textit{Chirality} \textbf{1992}, \textit{4}, 338--340}.
	
	\bibitem{Hutt1996} Hutt, A. J.; Tan, S. C. Drug chirality and its clinical significance. \href{https://doi.org/10.2165/00003495-199600525-00003}{\textit{Drugs} \textbf{1996}, \textit{52}, 1--12}.
	
	\bibitem{Smith2009} Smith, S.~W. Chiral toxicology: it's the same thing...only different. \href{https://doi.org/10.1093/toxsci/kfp097}{\textit{Toxicol. Sci.} \textbf{2009}, \textit{110}, 4--30}.
	
	\bibitem{Naaman2019} Naaman, R.; Paltiel, Y.; Waldeck, D.~H. Chiral molecules and the electron spin. \href{https://doi.org/10.1038/s41570-019-0087-1}{\textit{Nat. Rev. Chem.} \textbf{2019}, \textit{3}, 250--260}.
	
	\bibitem{Lodahl2017} Lodahl, P.; Mahmoodian, S.; Stobbe, S.; Rauschenbeutel, A.; Schneeweiss, P.; Volz, J.; Pichler, H.; Zoller, P. Chiral quantum optics. \href{https://doi.org/10.1038/nature21037}{\textit{Nature} \textbf{2017}, \textit{541}, 473--480}.
	
	\bibitem{Gohler2011} G\"ohler, B.; Hamelbeck, V.; Markus, T.~Z.; Kettner, M.; Hanne, G.~F.; Vager, Z.; Naaman, R.; Zacharias, H. Spin selectivity in electron transmission through self-assembled monolayers of double-stranded DNA. \href{https://doi.org/10.1126/science.1199339}{\textit{Science} \textbf{2011}, \textit{331}, 894--897}.
	
	\bibitem{Zhu2018} Zhu, H.; Yi, J.; Li, M.-Y.; Xiao, J.; Zhang, L.; Yang, C.-W.; Kaindl, R.~A.; Li, L.-J.; Wang, Y.; Zhang, X. Observation of chiral phonons. \href{https://doi.org/10.1126/science.aar2711}{\textit{Science} \textbf{2018}, \textit{359}, 579--582}.
	
	\bibitem{Cameron2012} Cameron, R.~P.; Barnett, S.~M.; Yao, A.~M. Optical helicity, optical spin and related quantities in electromagnetic theory. \href{https://doi.org/10.1088/1367-2630/14/5/053050}{\textit{New J. Phys.} \textbf{2012}, \textit{14}, 053050}.
	
	\bibitem{Alpeggiani2019} Alpeggiani, F.; Bliokh, K.~Y.; Nori, F.; Kuipers, L. Electromagnetic helicity in complex media. \href{https://doi.org/10.1103/PhysRevLett.120.243605}{\textit{Phys. Rev. Lett.} \textbf{2018}, \textit{120}, 243605}.
	
	\bibitem{Tang2010} Tang, Y.; Cohen, A.~E. Optical chirality and its interaction with matter. \href{https://doi.org/10.1103/PhysRevLett.104.163901}{\textit{Phys. Rev. Lett.} \textbf{2010}, \textit{104}, 163901}.
	
	\bibitem{Bliokh2011} Bliokh, K.~Y.; Nori, F. Characterizing optical chirality. \href{https://doi.org/10.1103/PhysRevA.83.021803}{\textit{Phys. Rev. A} \textbf{2011}, \textit{83}, 021803(R)}.
	
	\bibitem{Tang2011} Tang, Y.; Cohen, A.~E. Enhanced enantioselectivity in excitation of chiral molecules by superchiral light. \href{https://doi.org/10.1126/science.1202817}{\textit{Science} \textbf{2011}, \textit{332}, 333--336}.
	
	\bibitem{Berova} Berova, N.; Nakanishi, K.; Woody, R.~W. \textit{Circular Dichroism: Principles and Applications}; Wiley-VCH: New York, 2000.
	
	\bibitem{Barron} Barron, L.~D. \textit{Molecular Light Scattering and Optical Activity}; Cambridge Univ. Press: Cambridge, 2004.
	
	\bibitem{Hassey2006} Hassey, R.; Swain, E.~J.; Hammer, N.~I.; Venkataraman, D.; Barnes, M.~D. Probing the chiroptical response of a single molecule. \href{https://doi.org/10.1126/science.1134231}{\textit{Science} \textbf{2006}, \textit{314}, 1437--1439}.
	
	\bibitem{Hendry2010} Hendry, E.; Carpy, T.; Johnston, J.; Popland, M.; Mikhaylovskiy, R.~V.; Lapthorn, A.~J.; Kelly, S.~M.; Barron, L.~D.; Gadegaard, N.; Kadodwala, M. Ultrasensitive detection and characterization of biomolecules using superchiral fields. \href{https://doi.org/10.1038/nnano.2010.209}{\textit{Nat. Nanotechnol.} \textbf{2010}, \textit{5}, 783--787}.

	\bibitem{Rhee2013} Rhee, H.; Choi, J.~S.; Starling, D.~J., Howelld, J.~C.; Cho, M. Amplifications in chiroptical spectroscopy, optical enantioselectivity, and weak value measurement. \href{https://doi.org/10.1039/C3SC51255J}{\textit{Chem. Sci.} \textbf{2013}, \textit{4}, 4107--4114}.
	
	\bibitem{VazquezLozano2018} V\'azquez-Lozano, J.~E.; Mart\'inez, A. Optical chirality in dispersive and lossy media. \href{https://doi.org/10.1103/PhysRevLett.121.043901}{\textit{Phys. Rev. Lett.} \textbf{2018}, \textit{212}, 043901}.
	
	\bibitem{Schaferling} Sch\"aferling, M. \textit{Chiral Nanophotonics: Chiral Optical Properties of Plasmonic Systems}; Springer: Berlin, 2017.
	
	\bibitem{Schaferling2012} Sch\"aferling, M.; Dregely, D.; Hentschel, M.; Giessen, H. Tailoring enhanced optical chirality: Design principles for chiral plasmonic nanostructures. \href{https://doi.org/10.1103/PhysRevX.2.031010}{\textit{Phys. Rev. X} \textbf{2012}, \textit{2}, 031010}.
	
	\bibitem{Barr2018} Barr, L.~E.; Horsley, S.~A.~R.; Hooper, I.~R.; Eager, J.~K.; Gallagher, C.~P.; Hornett, S.~M.; Hibbins, A.~P.; Hendry, E. Investigating the nature of chiral near-field interactions. \href{https://doi.org/10.1103/PhysRevB.97.155418}{\textit{Phys. Rev. B} \textbf{2018}, \textit{97}, 155418}.
	
	\bibitem{Collins2017} Collins, J.~T.; Kuppe, C.; Hooper, D.~C.; Sibilia, C.; Centini, M.; Valev, V.~K. Chirality and chiroptical effects in metal nanostructures: Fundamentals and current trends. \href{https://doi.org/10.1002/adom.201700182}{\textit{Adv. Opt. Mater.} \textbf{2017}, \textit{5}, 1700182}.
	
	\bibitem{Hentschel2017} Hentschel, M.; Sch\"aferling, M.; Duan, X.; Giessen, H.; Liu, N. Chiral plasmonics. \href{https://doi.org/10.1126/sciadv.1602735}{\textit{Sci. Adv.} \textbf{2017}, \textit{3}, e1602735}.
		
	\bibitem{Govorov2010} Govorov, A.~O.; Fan, Z.; Hernandez, P.; Slocik, J.~M.; Naik, R.~R. Theory of circular dichroism of nanomaterials comprising chiral molecules and nanocrystals: Plasmon enhancement, dipole interactions, and dielectric effects. \href{https://doi.org/10.1021/nl100010v}{\textit{Nano Lett.} \textbf{2010}, \textit{10}, 1374--1382}.
	
	\bibitem{Zhao2017} Zhao, Y.; Askarpour, A.~N.; Sun, L.; Shi, J.; Li, X.; Al\`u, A. Chirality detection of enantiomers using twisted optical metamaterials. \href{https://doi.org/10.1038/ncomms14180}{\textit{Nat. Commun.} \textbf{2017}, \textit{8}, 14180}.
	
	\bibitem{Kang2017} Kang, L.; Ren, Q.; Werner, D.~H. Leveraging superchiral light for manipulation of optical chirality in the near-field of plasmonic metamaterials. \href{https://doi.org/10.1021/acsphotonics.7b00057}{\textit{ACS Photonics} \textbf{2017}, \textit{4}, 1298--1305}.
	
	\bibitem{Kramer2017} Kramer, C.; Sch\"aferling, M.; Weiss, T.; Giessen, H.; Brixner, T. Analytic optimization of near-field optical chirality enhancement. \href{https://doi.org/10.1021/acsphotonics.6b00887}{\textit{ACS Photonics} \textbf{2017}, \textit{4}, 396--406}.
	
	\bibitem{Hendry2012} Hendry, E.; Mikhaylovskiy, R.~V.; Barron, L.~D.; Kadodwala, M.; Davis, T.~J. Chiral electromagnetic fields generated by arrays of nanoslits. \href{https://doi.org/10.1021/nl3012787}{\textit{Nano Lett.} \textbf{2012}, \textit{12}, 3640--3644}.
	
	\bibitem{Meinzer2013} Meinzer, N.; Hendry, E.; Barnes, W.~L. Probing the chiral nature of electromagnetic fields surrounding plasmonic nanostructures. \href{https://doi.org/10.1103/PhysRevB.88.041407}{\textit{Phys. Rev. B} \textbf{2013}, \textit{88}, 041407(R)}.
	
	\bibitem{Nesterov2016} Nesterov, M.~L.; Yin, X.; Sch\"aferling, M.; Giessen, H.; Weiss, T. The role of plasmon-generated near fields for enhanced circular dichroism spectroscopy. \href{https://doi.org/10.1021/acsphotonics.5b00637}{\textit{ACS Photonics} \textbf{2016}, \textit{3}, 578--583}.
	
	\bibitem{Mohammadi2018} Mohammadi, E.; Tsakmakidis, K.~L.; Askarpour, A.~N.; Dehkhoda, P.; Tavakoli, A.; Altug, H. Nanophotonic platforms for enhanced chiral sensing. \href{https://doi.org/10.1021/acsphotonics.8b00270}{\textit{ACS Photonics} \textbf{2018}, \textit{5}, 2669--2675}.
	
	\bibitem{Pellegrini2018} Pellegrini, G.; Finazzi, M.; Celebrano, M.; Du\`o, L.; Biagioni, P. Chiral surface waves for enhanced circular dichroism. \href{https://doi.org/10.1103/PhysRevB.95.241402}{\textit{Phys. Rev. B} \textbf{2017}, \textit{95}, 241402(R)}.
	
	\bibitem{Solomon2019} Solomon, M.~L.; Hu, J.; Lawrence, M.; Garc\'ia-Etxarri, A.; Dionne, J.~A. Enantiospecific optical enhancement of chiral sensing and separation with dielectric metasurfaces. \href{https://doi.org/10.1021/acsphotonics.8b01365}{\textit{ACS Photonics} \textbf{2019}, \textit{6}, 43--49}.
	
	\bibitem{Graf2019} Graf, F.; Feis, J.; Garcia-Santiago, X.; Wegener, M.; Rockstuhl, C.; Fernandez-Corbaton, I. Achiral, helicity preserving, and resonant structures for enhanced sensing of chiral molecules. \href{https://doi.org/10.1021/acsphotonics.8b01454}{\textit{ACS Photonics} \textbf{2019}, \textit{6}, 482--491}.
	
	\bibitem{Mohammadi2019} Mohammadi, E.; Tavakoli, A.; Dehkhoda, P.; Jahani, Y.; Tsakmakidis, K.~L.; Tittl, A.; Altug, H. Accessible superchiral near-fields driven by tailored electric and magnetic resonances in all-dielectric nanostructures. \href{https://doi.org/10.1021/acsphotonics.8b01767}{\textit{ACS Photonics} \textbf{2019}, \textit{6}, 1939--1946}.
	
	\bibitem{Zhao2019} Zhao, X.; Reinhard, B.~M. Switchable chiroptical hot-spots in silicon nanodisk dimers. \href{https://doi.org/10.1021/acsphotonics.9b00388}{\textit{ACS Photonics} \textbf{2019}, \textit{6}, 1981--1989}.
	
	\bibitem{Estevez2012} Estevez, M.~C.; Alvarez, M.; Lechuga, L.~M. Integrated optical devices for lab-on-a-chip biosensing applications. \href{https://doi.org/10.1002/lpor.201100025}{\textit{Laser Photon. Rev.} \textbf{2012}, \textit{6}, 463--487}.
	
	\bibitem{Nie2017} Nie, X.; Ryckeboer, E.; Roelkens, G.; Baets, R. CMOS-compatible broadband co-propagative stationary Fourier transform spectrometer integrated on a silicon nitride photonics platform. \href{https://doi.org/10.1364/OE.25.00A409}{\textit{Opt. Express} \textbf{2017}, \textit{25}, A409--A418}.
	
	\bibitem{Petersen2014} Petersen, J.; Volz, J.; Rauschenbeutel, A. Chiral nanophotonic waveguide interface based on spin-orbit interaction of light. \href{https://doi.org/10.1126/science.1257671}{\textit{Science} \textbf{2014}, \textit{346}, 67--71}.
	
	\bibitem{Coles2016} Coles, R.~J.; Price, D.~M.; Dixon, J.~E.; Royall, B.; Clarke, E.; Kok, P.; Skolnick, M.~S.; Fox, A.~M.; Makhonin, M.~N. Chirality of nanophotonic waveguide with embedded quantum emitter for unidirectional spin transfer. \href{https://doi.org/10.1038/ncomms11183}{\textit{Nat. Commun.} \textbf{2016}, \textit{7}, 11183}.
	
	\bibitem{Gong2018} Gong, S.-H.; Alpeggiani, F.; Sciacca, B.; Garnett, E.~C.; Kuipers, L. Nanoscale chiral valley-photon interface through optical spin-orbit coupling. \href{https://doi.org/10.1126/science.aan8010}{\textit{Science} \textbf{2018}, \textit{359}, 443--447}.
	
	\bibitem{Picardi2018} Picardi, M.~F.; Bliokh, K.~Y.; Rodr\'iguez-Fortu\~no, F.~J.; Alpeggiani, F.; Nori, F. Angular momenta, helicity, and other properties of dielectric-fiber and metallic-wire modes. \href{https://doi.org/10.1364/OPTICA.5.001016}{\textit{Optica} \textbf{2018}, \textit{5}, 1016--1026}.
	
	\bibitem{Kien2017} Le Kien, F.; Busch, T.; Truong, V.~G.; Chormaic, S.~N. Higher-order modes of vacuum-clad ultrathin optical fibers. \href{https://doi.org/10.1103/PhysRevA.96.023835}{\textit{Phys. Rev. A} \textbf{2017}, \textit{96}, 023835}.
	
	\bibitem{Abujetas2018} Abujetas, D.~R.; S\'anchez-Gil, J.~A. Spin angular momentum in planar and cylindrical waveguides induced by transverse confinement and intrinsic helicity of guided light. \href{https://arxiv.org/abs/1809.02406}{\textit{arXiv:1809.02406} \textbf{2018}}.
	
	\bibitem{Saleh} Saleh, B.~E.~A.; Teich, M.~C. \textit{Fundamentals of Photonics}; Wiley-Interscience: New York, 2007.
	
	\bibitem{Bliokh2012} Bliokh, K.~Y.; Nori, F. Transverse spin of a surface polariton. \href{https://doi.org/10.1103/PhysRevA.85.061801}{\textit{Phys. Rev. A} \textbf{2012}, \textit{85}, 061801(R)}.
	
	\bibitem{Alizadeh2015a} Alizadeh, M.; Reinhard, B.~M. Enhanced optical chirality through locally excited surface plasmon polaritons. \href{https://doi.org/10.1021/acsphotonics.5b00151}{\textit{ACS Photonics} \textbf{2015}, \textit{2}, 942--949}.
	
	\bibitem{Nechayev2019} Nechayev, S.; Barczyk, R.; Mick, U.; Banzer, P. Substrate-induced chirality in an individual nanostructure. \href{https://doi.org/10.1021/acsphotonics.9b00748}{\textit{ACS Photonics} \textbf{2019}, \textit{6}, 1876--1881}.
	
	\bibitem{Petronijevic2019} Petronijevic, E.; Sibilia, C. Enhanced near-field chirality in periodic arrays of Si nanowires for chiral sensing. \href{https://doi.org/10.3390/molecules24050853}{\textit{Molecules} \textbf{2019}, \textit{24}, 853}.
	
	\bibitem{RomeroGarcia2013} Romero-Garc\'ia, S.; Merget, F.; Zhong, F.; Finkelstein, H.; Witzens, J. Silicon nitride CMOS-compatible platform for integrated photonics applications at visible wavelengths. \href{https://doi.org/10.1364/OE.21.014036}{\textit{Opt. Express} \textbf{2013}, \textit{21}, 14036--14046}.
	
	\bibitem{EspinosaSoria2016a} Espinosa-Soria, A.; Mart\'inez, A. Transverse spin and spin-orbit coupling in silicon waveguides. \href{https://doi.org/10.1109/LPT.2016.2553841}{\textit{IEEE Photonics Technol. Lett.} \textbf{2016}, \textit{28}, 1561--1564}.
	
	\bibitem{Almeida2004} Almeida, V.~R.; Xu, Q.; Barrios, C.~A.; Lipson, M. Guiding and confining light in void nanostructure. \href{https://doi.org/10.1364/OL.29.001209}{\textit{Opt. Lett.} \textbf{2004}, \textit{29}, 1209--1211}.
	
	\bibitem{Barrios2007} Barrios, C.~A.; Gylfason, K.~B.; S\'anchez, B.; Griol, A.; Sohlstr\"om, H.; Holgado, M.; Casquel, R. Slot-waveguide biochemical sensor. \href{https://doi.org/10.1364/OL.32.003080}{\textit{Opt. Lett.} \textbf{2007}, \textit{32}, 3080--3082}.
	
	\bibitem{Gansel2009} Gansel, J.~K.; Thiel, M.; Rill, M.~S.; Decker, M.; Bade, K.; Saile, V.; von Freymann, G.; Linden, S.; Wegener, M. Gold helix photonic metamaterial as broadband circular polarizer. \href{https://doi.org/10.1126/science.1177031}{\textit{Science} \textbf{2009}, \textit{325}, 1513--1515}.

	\bibitem{Schaferling2014} Sch\"aferling, M.; Yin, X.; Engheta, N.; Giessen, H. Helical plasmonic nanostructures as prototypical chiral near-field sources. \href{https://doi.org/10.1021/ph5000743}{\textit{ACS Photonics} \textbf{2014}, \textit{1}, 530--537}.
	
	\bibitem{Esposito2015} Esposito, M.; Tasco, V.; Cuscun\`a, M.; Todisco, F.; Benedetti, A.; Tarantini, I.; De Giorgi, M.; Sanvitto, D.; Passaseo, A. Nanoscale 3D chiral plasmonic helices with circular dichroism at visible frequencies. \href{https://doi.org/10.1021/ph500318p}{\textit{ACS Photonics} \textbf{2015}, \textit{2}, 105--114}.
	
	\bibitem{Ji2016} Ji, R.; Wang, S.-W.; Liu, X.; Guo, H.; Lu, W. Hybrid helix metamaterials for giant and ultrawide circular dichroism. \href{https://doi.org/10.1021/acsphotonics.6b00575}{\textit{ACS Photonics} \textbf{2016}, \textit{3}, 2368--2374}.
	
	\bibitem{Wozniak2018} Wo\`zniak, P.; De Leon, I.; Ho\"oflich, K.; Haverkamp, C.; Christiansen, S.; Leuchs, G.; Banzer, P. Chiroptical response of a single plasmonic nanohelix. \href{https://doi.org/10.1364/OE.26.019275}{\textit{Opt. Express} \textbf{2018}, \textit{26}, 19275--19293}.
	
	\bibitem{Hoflich2019} H\"oflich, K.; Feichtner, T.; Hansj\"urgen, E.; Haverkamp, C.; Kollmann, H.; Lienau, C.; Silies, M. Resonant behavior of a single plasmonic helix. \href{https://doi.org/10.1364/OPTICA.6.001098}{\textit{Optica} \textbf{2019}, \textit{6}, 1098--1105}.
	
	\bibitem{Johnson1972} Johnson, P.~B.; Christy, R.~W. Optical constants of the noble metals. \href{https://doi.org/10.1103/PhysRevB.6.4370}{\textit{Phys. Rev. B} \textbf{1972}, \textit{6}, 4370--4379}.
	
	\bibitem{EspinosaSoria2016b} Espinosa-Soria, A.; Griol, A.; Mart\'inez, A. Experimental measurement of plasmonic nanostructures embedded in silicon waveguide gaps. \href{https://doi.org/10.1364/OE.24.009592}{\textit{Opt. Express} \textbf{2016}, \textit{24}, 9592--9601}.
	
	\bibitem{EspinosaSoria2018} Espinosa-Soria, A.; Pinilla-Cienfuegos, E.; D\'iaz-Fern\'andez, F.~J.; Griol, A.; Mart\'i, J.; Mart\'inez, A. Coherent control of a plasmonic nanoantenna integrated on a silicon chip. \href{https://doi.org/10.1021/acsphotonics.8b00447}{\textit{ACS Photonics} \textbf{2018}, \textit{5}, 2712--2717}.
	
	\bibitem{Filippov1990} Filippov, V.~N.; Kotov, O.~I.; Nikolayev; V.~M. Measurement of polarisation beat length in single-mode optical fibres with a polarisation modulator. \href{https://doi.org/10.1049/el:19900431}{\textit{Electron. Lett.} \textbf{1990}, \textit{26}, 658--660}.
	
	\bibitem{Zhang2019} Zhang, Q.; Hernandez, T.; Smith, K.~W.; Jebeli, S.~A.~H.; Dai, A.~X.; Warning, L.; Baiyasi, R.; McCarthy, L.~A.; Guo, H.; Chen, D.-H.; Dionne, J.~A.; Landes, C.~F.; Link, S. Unraveling the origin of chirality from plasmonic nanoparticle-protein complexes. \href{https://doi.org/10.1126/science.aax5415}{\textit{Science} \textbf{2019}, \textit{365}, 1475--1478}.
	
	\bibitem{Schaferling2016} Sch\"aferling, M.; Engheta, N.; Giessen, H.; Weiss, T. Reducing the complexity: Enantioselective chiral near-fields by diagonal slit and mirror configuration. \href{https://doi.org/10.1021/acsphotonics.6b00147}{\textit{ACS Photonics} \textbf{2016}, \textit{3}, 1076--1084}.
	
	\bibitem{Meca2017} Garc\'ia-Meca, C.; Lechago, S.; Brimont, A.; Griol, A.; Mas, S.; S\'anchez, L.; Bellieres, L.; Losilla, N.~S.; Mar\'i, J. On-chip wireless silicon photonics: from reconfigurable interconnects to lab-on-chip devices. \href{https://doi.org/10.1038/lsa.2017.53}{\textit{Light: Sci. Appl.} \textbf{2017}, \textit{6}, e17053}.
\end{thebibliography}
\end{document}